%% file: main.tex
\documentclass{llncs}
\usepackage[T1]{fontenc}
\usepackage[colorlinks=true, linkcolor=blue, citecolor=blue, urlcolor=blue]{hyperref}

\usepackage{graphicx}
\usepackage{orcidlink}

\usepackage{xcolor}

\graphicspath{ {./fig/}  } 

\begin{document}
\title{SSRLBot: Designing and Developing a Large Language Model-based Agent using Socially Shared Regulated Learning}
%
%
\author{
Xiaoshan Huang\orcidlink{0000-0002-2853-7219}\thanks{Equal contributions.} \and
Jie Gao\orcidlink{0000-0002-6933-950X}$^{\ast}$ \and
Haolun Wu\orcidlink{0000-0001-6255-1535}
}

\institute{
McGill University, Montreal, Quebec, Canada\\
\email{\{xiaoshan.huang, jie.gao3, haolun.wu\}@mail.mcgill.ca}
}

%
\maketitle              
\begin{abstract}
Large language model (LLM)--based agents have emerged as pivotal tools in assisting human experts across various fields by transforming complex tasks into more efficient workflows and providing actionable stakeholder insights. Despite their potential, the application of LLM-based agents for medical education remains underexplored. The study aims to assist in evaluating the students' process and outcomes on medical case diagnosis and discussion while incorporating the theoretical framework of Socially Shared Regulation of Learning (SSRL)  to assess student performance. SSRL emphasizes metacognitive, cognitive, motivational, and emotional interactions, highlighting the collaborative management of learning processes to improve decision-making outcomes. Grounded in SSRL theory, this tool paper introduces SSRLBot, an LLM-based agent designed to enable team members to reflect on their diagnostic performance and the key SSRL skills that foster team success. SSRLBot's core functions include summarizing dialogue content, analyzing participants' SSRL skills, and evaluating students' diagnostic results. Meanwhile, we evaluated SSRLBot through diagnostic conversation data collected from six groups (12 participants, 1926 conversational turns). Results showed that SSRLBot can deliver detailed, theory-aligned evaluations, link specific behaviors to SSRL dimensions, and offer actionable recommendations for improving teamwork. The findings address a critical gap in medical education, advancing the application of LLM agents to enhance team-based decision-making and collaboration in high-stakes environments.

\keywords{Generative Artificial Intelligence \and Large Language Models \and Socially Shared Regulation \and Medical Education \and Conversational Agents.}
\end{abstract}
\input{01_introduction}
\input{02_related_work}
\input{03_SSRLBot}

\input{04_CaseStudy}

\input{05_Conclusion}

\newpage
\bibliographystyle{splncs04}
\bibliography{mybibliography}

\end{document}

%% file: 01_introduction.tex
\section{Introduction}
Team actions—decisions and operations carried out collaboratively—are essential for achieving success. However, team action alone may not guarantee success without the joint efforts of team members in adopting appropriate strategies, operations, and regulations ~\cite{salas2005}. This is especially true in high-stakes decision-making contexts, such as clinical reasoning, where a single decision can lead to significantly different outcomes ~\cite{djulbegovic2017progress}. The effectiveness of team decision-making is closely tied to the Socially Shared Regulation of Learning (SSRL, ~\cite{panadero2015socially}). This framework emphasizes the collaborative regulation of learning processes as key to improving decision-making outcomes. In medical teamwork, team members’ interactions through conversation play a pivotal role in their diagnostic outcome, as they could either direct to the correct or an opposite result ~\cite{ball2015improving}. Medical simulation has been pivotal for medical professionals to practice diagnostic skills and paves the way for understanding how teamwork works in this field ~\cite{davis2021health,rosen2008promoting}.

In this study, we designed and developed a large language model (LLM)-based agent, SSRLBot. SSRLBot can evaluate students' performance in medical case diagnostics and conversations, judge the diagnosis results, and identify students' SSRL skills in their conversations. Grounded on the theory of SSRL, SSRLBot enables students to reflect on their diagnostic performance as a team and the key SSRL skills used to foster team diagnostic success in medical simulation. Additionally, we used six groups' diagnostic conversation data, including 12 participants and 1926 conversational turns, to evaluate this LLM-based agent.

%% file: 02_related_work.tex
\section{Background and Related Work}
\subsection{SSRL in Team Work}
Socially shared regulation is essential in teamwork. It involves team members' mutual efforts in regulating team performance through effective interactions. Socially shared regulated learning reflects learning strategies in educational contexts where two or more learners are involved. Studies in collaborative learning have found that groups that perform better show more SSRL strategies in teamwork ~\cite{malmberg2015promoting}. However, the mechanism of SSRL is complex and hard to be captured, as dynamic interpersonal interactions occur where team members are influencing each other. SSRL guides interactions in teamwork from four dimensions, namely meta-cognitive, socio-cognitive, socio-motivational, and socio-emotional aspects. The meta-cognitive aspect of interactions (short in meta-cognitive interaction in the following content) refers to conversations about the team's cognitive progress ~\cite{haataja2022individuals,nguyen2023examining}. Cognitive interaction includes conversations about the team's actual cognitive activities, such as planning, performing, and reflecting on the task and team performance ~\cite{nguyen2023examining}. Meta-cognitive interaction to teamwork is like an umbrella to an outdoor walking with a weather forecast; it prevents rains when necessary and can be folded when the sun comes out. Socio-emotional interaction, or social-emotional interaction, is team members' efforts to maintain cohesive and respectful social interaction through team dynamics ~\cite{jarvela2015enhancing}. Healthy and positive socio-emotional interaction enables team members to engage more effectively in cognitive interaction and leads to better team outcomes. Socio-motivational interaction is often discussed with socio-emotional interactions ~\cite{naykki2014socio,jarvela2015enhancing}, as it is pivotal to motivating team spirits and triggers team members to stay tuned and engaged in teamwork.

SSRL research often conducts discourse analysis in computer-supported collaborative learning (CSCL) contexts, as it allows for the identification of learners' situational social, cognitive, emotional, and motivational states through transcriptions of their dialogues or written texts~\cite{hernandez2019computer}. Integrating trace data from think-aloud protocols and CSCL logfiles represents a significant advancement in capturing learners' thought processes and actions, providing a more objective perspective on SSRL~\cite{azevedo2013using}. Most studies analyze verbal data through human coding, which, despite expert evaluation, remains labor-intensive and prone to bias. LLMs learn word usage patterns and apply them to natural language processing tasks~\cite{thirunavukarasu2023large}, offering a more efficient alternative that often matches or surpasses human performance in content understanding, idea generation, and decision-making.

\subsection{LLM Agent in Education}
LLM agents have recently debuted in educational contexts. For example, PBChat is designed to identify and provide solutions to students’ problem behaviors using teacher-parent conversations. LLMAgent-CK ~\cite{yang2024content} is designed as a multi-agent framework to identify middle-school teachers’ mathematical content knowledge learning goals from their responses. Other LLM-based agents are designed as virtual tutors or assistants in diverse subjects (e.g., Ruffle \& Riley: Biology ~\cite{schmucker2024ruffle}; Jill Watsons, course readings ~\cite{taneja2024jill}). Furthermore, other LLM-based agents have been designed for advancing educational research (e.g., Vizchat for visualizing multimodal learning analytics ~\cite{yan2024vizchat}). Despite previous applications, their potential for medical education remains unexplored. The SSRL theory-driven design could provide valuable insights into teamwork evaluation. Notably, SSRLBot is the first LLM-based agent designed to assess medical teams’ SSRL interactions and their impact on diagnostic performance. 

%% file: 03_SSRLBot.tex
\section{SSRLBot: System Architecture} 

\begin{figure}[t]
    \centering
    \includegraphics[width=1.0\linewidth]{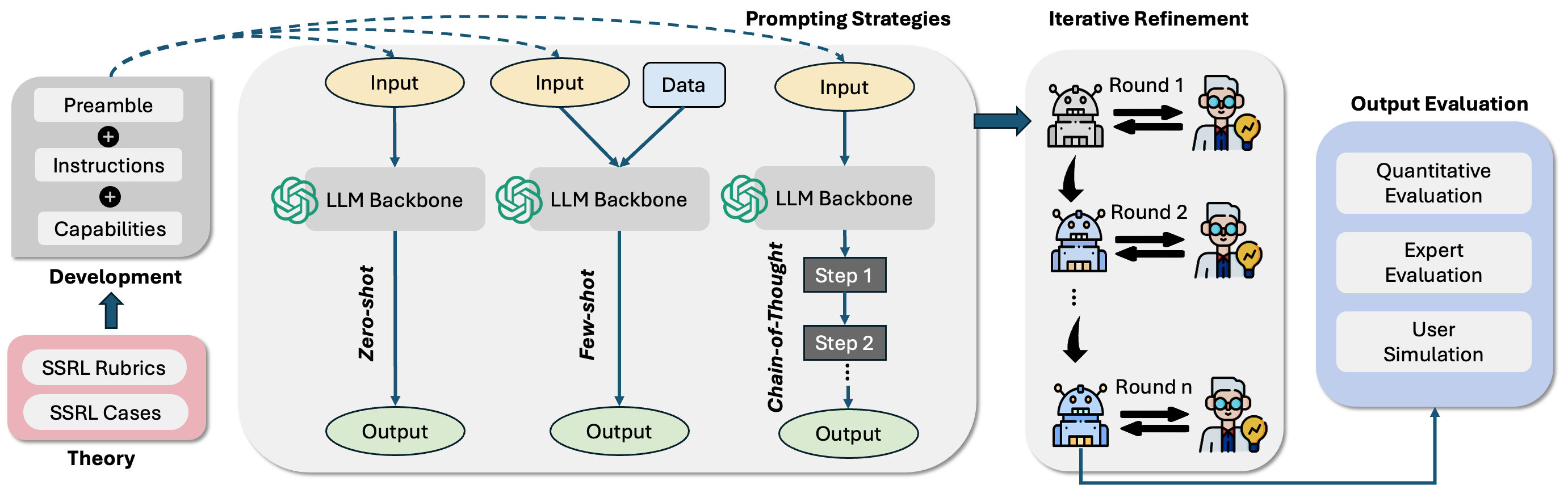}
    \caption{The Workflow of the System Architecture for SSRLBot.}
    \label{fig:prototype}
\end{figure}

Figure~\ref{fig:prototype} shows the entire diagram from data collection to output evaluation, consisting of three key sections: 1) the theory-based development section, 2) the tuning section, and 3) the evaluation section. The architecture of SSRLBot integrates the GPT model (e.g., GPT-4), instructions, and SSRL rubrics. SSRLBot aims to improve the progress of data processing and optimize the quality of the data, reducing human-caused errors in data annotating and diagnosing. In addition, SSRLBot can provide actionable insights for researchers to improve their abilities in assessing and developing SSRL skills in educational, clinical, and collaborative learning environments.

\subsection{Design}

Since OpenAI launched the GPT store, GPT apps with different purposes have amplified the functionality of ChatGPT, making ChatGPT fully applicable to meet the needs of various users. GPT app configuration consists of four core components, including instructions, knowledge, capabilities, and actions. Instructions refer to development details of the GPT app configuration, such as the app’s name, description, logo, instruction description, and conversation starters. Knowledge refers to the external knowledge files that can help improve response accuracy, fairness, and user experience~\cite{zhang2024first}. Users can download these files if the code interpreter is enabled. Capabilities refer to web search, canvas, DALL·E Image Generation, and code interpreter and data analysis. Actions refer to make a third-party APIs to the GPT app. SSRLBot focuses on the instructions and capabilities of the GPT app framework. Furthermore, the SSRL framework is used to support specific functions under the instructions.

\subsection{Functional Realization}

\begin{figure}[t]
    \centering
    \includegraphics[width=0.9\linewidth]{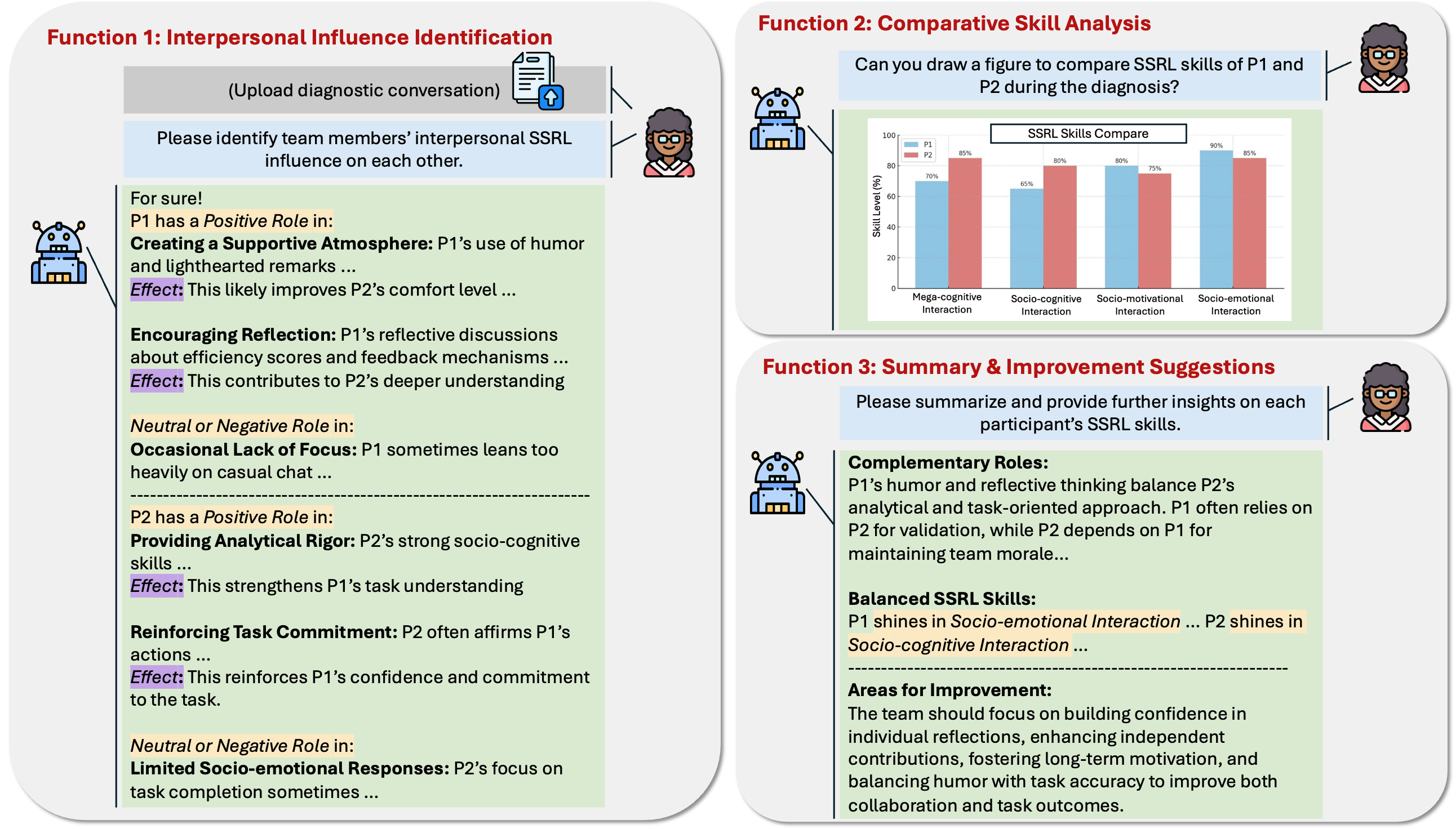}
    \caption{The Demonstration of SSRLBot Functions.}
    \label{fig:demonstration}
\end{figure}

The instruction description consists of a preamble and function descriptions. The preamble's setting helps SSRLBot clearly understand its role and continue playing this role in a scene for dialogue. We set the SSRLBot as a professional researcher with enough SSRL knowledge and health knowledge, be able to analyze and evaluate data. Moreover, we developed three core functions in SSRLBot: (1) identify the participant's interpersonal influence on each, (2) compare the SSRL skills, and (3) summarize and provide suggestions on participants. Figure~\ref{fig:demonstration} illustrates functions of SSRLBot. To enable SSRLBot to identify corresponding SSRL skills effectively and accurately, we developed an SSRL framework and integrated it into the system's preamble. This framework consists of three SSRL skill layers. The first layer contains five core SSRL skills: meta-cognitive interaction, socio-cognitive interaction, socio-motivational interaction, socio-emotional interaction, and task execution. Each core SSRL skill also contains the second and third SSRL skill layers. Definitions and examples were provided for each skill to help the system make more accurate judgments.

Although the SSRLBot has been set with three core functions to achieve the goal, the output is not guaranteed to be completely accurate. Tuning plays a key role in building the SSRLBot and enhancing dialogue competencies and accuracy. The tuning section in this study focuses primarily on two parts: instruction and prompt. Instruction tuning helps provide feedback to system development and optimize instructions and preamble. Prompt tuning helps improve input quality. Previous studies~\cite{diao2023active,song2023comprehensive,zhang2023multimodal} have demonstrated that using more appropriate prompts can improve output quality and mitigate hallucinations. This study conducted four rounds of iterative refinement. The first two rounds focused on functional completeness. The third round was mainly adjusted for output accuracy and quality. The fourth round focused on prompt improvement.

%% file: 04_CaseStudy.tex
\section{Illustrative Case Study}

In this section, we evaluated SSRLBot using six groups of authentic conversational data, which were collected from 12 medical residents recruited from a university in Montreal, Canada. 

\textit{Data collection.} Participants were paired up and required to discuss a medical case through the BioWorld system. BioWorld is a computer learning system that can provide a hospital simulation for students to discuss diseases using related knowledge ~\cite{lajoie2001constructing}. The duration of each conversation was around 30 minutes. The number of conversational turn in each group was between 123 and 600. We conducted basic data cleaning on the collected data.  For example, we added a speaker  (e.g., P1 for participant 1, P2 for participant 2, R for researcher) and the corresponding SSRL skill on each conversational turn. In addition, we evaluated the diagnostic results of each group with the standard diagnostic answer to determine whether each group's results were correct.  

\textit{Evaluation.} Each group's conversational data was saved as a case in a single file. In this study, we mainly evaluated the diagnostic performance of SSRLBot. We requested SSRLBot summarize the conversation content for each case and evaluate the diagnostic results. The responses provided by SSRLBot primarily include three sections: a summary of the conversation, an evaluation of diagnostic accuracy, and a conclusion. This agent enhanced the precision of its outcomes by extracting the keywords and critical conversational content as evidence. Researchers compared the SSRLBot's results and the standard results and found that SSRLBot demonstrated 100\% diagnostic accuracy.

\textit{Comparison.} We mainly compare the performance of SSRLBot with other LLMs, including ChatGPT-3.5, Deepseek-R1, and Gemini-1.5. All selected LLMs demonstrated their ability to understand the conversation's goal, identify key information, and generate a general report based on the given content. However, ChatGPT 3.5 relied less on contextualized descriptions for each participant, limiting its usefulness for individuals aiming to improve their teamwork skills. DeepSeek-R1 provided a more thorough evaluation by blending positive and critical feedback on participants' SSRL skills and their impact on diagnostic performance. However, its theoretical alignment could be improved by explicitly linking behaviors to SSRL dimensions and offering more personalized feedback. Gemini-1.5 provided contextualized descriptions; for example, it acknowledged humor's role within the team and aligned with SSRL theory. However, its analysis lacked actionable feedback for human-centered reflection. In contrast, SSRLBot can follow the SSRL framework to better identify the specific SSRL skills from the conversation while highlighting specific examples. Additionally, it provided a comprehensive analysis by linking evidence from the conversation to each participant's SSRL skills, guiding researchers on which specific SSRL skill layers to adapt for improved team decision-making.

%% file: 05_Conclusion.tex
\section{Conclusion and Future Work}
This paper introduced SSRLBot, an LLM-based agent for medical education practice grounded in the SSRL framework. SSRLBot aims to facilitate customized feedback for learners through evaluating their conversations in the collaborative learning environment. This theory-driven knowledge-empowering design enhances LLM functions that support teamwork enhancement through customized feedback for individuals and groups. Our case study demonstrates that SSRLBot's responses are highly accurate. The limitation of this study is the limited number of cases for evaluation. Although each case included extensive conversational data, it is important to conduct more cases to further evaluate the accuracy and effectiveness of this agent. Future research should expand the sample size and evaluate other functions of SSRLBot. In addition, future research can enhance SSRLBot by expanding its application beyond dyadic medical diagnosis to diverse contexts, enabling a more comprehensive evaluation of its theory-driven functions. Additionally, while researchers assessed its contex- tual functions, incorporating participant feedback would provide a more robust measure of its reliability and informativeness. Overall, the functionalities of this agent can be beneficial for medical education by helping assess learners' performance and SSRL skills and providing effective summarization and evaluation evidence for researchers.

%% file: main.bbl
\begin{thebibliography}{10}
\providecommand{\url}[1]{\texttt{#1}}
\providecommand{\urlprefix}{URL }
\providecommand{\doi}[1]{https://doi.org/#1}

\bibitem{azevedo2013using}
Azevedo, R., Harley, J., Trevors, G., Duffy, M., Feyzi-Behnagh, R., Bouchet, F., Landis, R.: Using trace data to examine the complex roles of cognitive, metacognitive, and emotional self-regulatory processes during learning with multi-agent systems. International handbook of metacognition and learning technologies pp. 427--449 (2013)

\bibitem{ball2015improving}
Ball, J.R., Miller, B.T., Balogh, E.P.: Improving diagnosis in health care  (2015)

\bibitem{davis2021health}
Davis, J., Zulkosky, K., Ruth-Sahd, L.A., Frank, E.M., Dommel, L., Minchhoff, D., Uhrich, K.: Health care professional students’ perceptions of teamwork and roles after an interprofessional critical care simulation. Dimensions of Critical Care Nursing  \textbf{40}(3),  174--185 (2021)

\bibitem{diao2023active}
Diao, S., Wang, P., Lin, Y., Pan, R., Liu, X., Zhang, T.: Active prompting with chain-of-thought for large language models. arXiv preprint arXiv:2302.12246  (2023)

\bibitem{djulbegovic2017progress}
Djulbegovic, B., Guyatt, G.H.: Progress in evidence-based medicine: a quarter century on. The lancet  \textbf{390}(10092),  415--423 (2017)

\bibitem{haataja2022individuals}
Haataja, E., Dindar, M., Malmberg, J., J{\"a}rvel{\"a}, S.: Individuals in a group: Metacognitive and regulatory predictors of learning achievement in collaborative learning. Learning and Individual Differences  \textbf{96},  102146 (2022)

\bibitem{hernandez2019computer}
Hern{\'a}ndez-Sell{\'e}s, N., Mu{\~n}oz-Carril, P.C., Gonz{\'a}lez-Sanmamed, M.: Computer-supported collaborative learning: An analysis of the relationship between interaction, emotional support and online collaborative tools. Computers \& Education  \textbf{138},  1--12 (2019)

\bibitem{jarvela2015enhancing}
J{\"a}rvel{\"a}, S., Kirschner, P.A., Panadero, E., Malmberg, J., Phielix, C., Jaspers, J., Koivuniemi, M., J{\"a}rvenoja, H.: Enhancing socially shared regulation in collaborative learning groups: Designing for cscl regulation tools. Educational Technology Research and Development  \textbf{63},  125--142 (2015)

\bibitem{lajoie2001constructing}
Lajoie, S.P., Guerrera, C., Munsie, S.D., Lavigne, N.C.: Constructing knowledge in the context of bioworld. Instructional Science  \textbf{29},  155--186 (2001)

\bibitem{malmberg2015promoting}
Malmberg, J., J{\"a}rvel{\"a}, S., J{\"a}rvenoja, H., Panadero, E.: Promoting socially shared regulation of learning in cscl: Progress of socially shared regulation among high-and low-performing groups. Computers in Human Behavior  \textbf{52},  562--572 (2015)

\bibitem{naykki2014socio}
N{\"a}ykki, P., J{\"a}rvel{\"a}, S., Kirschner, P.A., J{\"a}rvenoja, H.: Socio-emotional conflict in collaborative learning—a process-oriented case study in a higher education context. International Journal of Educational Research  \textbf{68},  1--14 (2014)

\bibitem{nguyen2023examining}
Nguyen, A., J{\"a}rvel{\"a}, S., Ros{\'e}, C., J{\"a}rvenoja, H., Malmberg, J.: Examining socially shared regulation and shared physiological arousal events with multimodal learning analytics. British Journal of Educational Technology  \textbf{54}(1),  293--312 (2023)

\bibitem{panadero2015socially}
Panadero, E., J{\"a}rvel{\"a}, S.: Socially shared regulation of learning: A review. European Psychologist  (2015)

\bibitem{rosen2008promoting}
Rosen, M.A., Salas, E., Wu, T.S., Silvestri, S., Lazzara, E.H., Lyons, R., Weaver, S.J., King, H.B.: Promoting teamwork: An event-based approach to simulation-based teamwork training for emergency medicine residents. Academic Emergency Medicine  \textbf{15}(11),  1190--1198 (2008)

\bibitem{salas2005}
Salas, E., Sims, D.E., Burke, C.S.: Is there a “big five” in teamwork? Small group research  \textbf{36}(5),  555--599 (2005)

\bibitem{schmucker2024ruffle}
Schmucker, R., Xia, M., Azaria, A., Mitchell, T.: Ruffle\&riley: Insights from designing and evaluating a large language model-based conversational tutoring system. In: International Conference on Artificial Intelligence in Education. pp. 75--90. Springer (2024)

\bibitem{song2023comprehensive}
Song, Y., Wang, T., Cai, P., Mondal, S.K., Sahoo, J.P.: A comprehensive survey of few-shot learning: Evolution, applications, challenges, and opportunities. ACM Computing Surveys  \textbf{55}(13s),  1--40 (2023)

\bibitem{taneja2024jill}
Taneja, K., Maiti, P., Kakar, S., Guruprasad, P., Rao, S., Goel, A.K.: Jill watson: A virtual teaching assistant powered by chatgpt. In: International Conference on Artificial Intelligence in Education. pp. 324--337. Springer (2024)

\bibitem{thirunavukarasu2023large}
Thirunavukarasu, A.J., Ting, D.S.J., Elangovan, K., Gutierrez, L., Tan, T.F., Ting, D.S.W.: Large language models in medicine. Nature medicine  \textbf{29}(8),  1930--1940 (2023)

\bibitem{yan2024vizchat}
Yan, L., Zhao, L., Echeverria, V., Jin, Y., Alfredo, R., Li, X., Ga{\v{s}}evi’c, D., Martinez-Maldonado, R.: Vizchat: enhancing learning analytics dashboards with contextualised explanations using multimodal generative ai chatbots. In: International Conference on Artificial Intelligence in Education. pp. 180--193. Springer (2024)

\bibitem{yang2024content}
Yang, K., Chu, Y., Darwin, T., Han, A., Li, H., Wen, H., Copur-Gencturk, Y., Tang, J., Liu, H.: Content knowledge identification with multi-agent large language models (llms). In: International Conference on Artificial Intelligence in Education. pp. 284--292. Springer (2024)

\bibitem{zhang2024first}
Zhang, Z., Zhang, L., Yuan, X., Zhang, A., Xu, M., Qian, F.: A first look at gpt apps: Landscape and vulnerability. arXiv preprint arXiv:2402.15105  (2024)

\bibitem{zhang2023multimodal}
Zhang, Z., Zhang, A., Li, M., Zhao, H., Karypis, G., Smola, A.: Multimodal chain-of-thought reasoning in language models. arXiv preprint arXiv:2302.00923  (2023)

\end{thebibliography}
